\documentclass[sigplan,screen,natbib,10pt,nonacm]{acmart}

\def\BibTeX{{\rm B\kern-.05em{\sc i\kern-.025em b}\kern-.08emT\kern-.1667em\lower.7ex\hbox{E}\kern-.125emX}}

\setcopyright{acmcopyright}
\copyrightyear{2019}
\acmYear{2019}

\usepackage[paperVersion]{dpupaper}

\newcommand{\reported}{%
\begin{tikzpicture}[baseline=0.1em,scale=0.9]
    \fill[rounded corners=0.8pt,fill=blue, fill opacity=0.4] (0,0) -- (0,0.21) -- (0.21,0.21) -- (0.21,0) -- (0,0);
    \fill[rounded corners=0.8pt,fill=red, fill opacity=0.75] (0.3,0.3) -- (0.09,0.3) -- (0.09,0.09) -- (0.3,0.09) -- (0.3,0.3);   

    \draw[semithick,rounded corners=0.8pt] (0,0) -- (0,0.21) -- (0.21,0.21) -- (0.21,0) -- (0,0);
    \draw[semithick,rounded corners=0.8pt] (0.3,0.3) -- (0.09,0.3) -- (0.09,0.09) -- (0.3,0.09) -- (0.3,0.3); 
\end{tikzpicture}\xspace}

\newcommand{\userfacing}{%
\begin{tikzpicture}[baseline=0.1em,scale=0.9]
    \fill[rounded corners=0.8pt,fill=blue, fill opacity=0.4] (0,0) -- (0,0.21) -- (0.21,0.21) -- (0.21,0) -- (0,0);
    \fill[rounded corners=0.7pt,fill=white] (0.09,0.21) -- (0.09,0.09) -- (0.21,0.09);
    \fill[rounded corners=0.7pt,fill=white] (0.21,0.09) -- (0.21,0.21) -- (0.09,0.21);

    \draw[semithick,rounded corners=0.8pt] (0,0) -- (0,0.21) -- (0.21,0.21) -- (0.21,0) -- (0,0);
    \draw[semithick,rounded corners=0.8pt] (0.3,0.3) -- (0.09,0.3) -- (0.09,0.09) -- (0.3,0.09) -- (0.3,0.3);
\end{tikzpicture}\xspace}

\newcommand{\partrep}{%
\begin{tikzpicture}[baseline=0.1em,scale=0.9]
    \fill[rounded corners=0.8pt,fill=blue, fill opacity=0.4] (0,0) -- (0,0.21) -- (0.21,0.21) -- (0.21,0) -- (0,0);
    \fill[rounded corners=0.7pt,fill=red, fill opacity=0.75] (0.09,0.21) -- (0.09,0.09) -- (0.21,0.09);
    \fill[rounded corners=0.7pt,fill=red, fill opacity=0.75] (0.21,0.09) -- (0.21,0.21) -- (0.09,0.21);

    \draw[semithick,rounded corners=0.8pt] (0,0) -- (0,0.21) -- (0.21,0.21) -- (0.21,0) -- (0,0);
    \draw[semithick,rounded corners=0.8pt] (0.3,0.3) -- (0.09,0.3) -- (0.09,0.09) -- (0.3,0.09) -- (0.3,0.3);
\end{tikzpicture}\xspace}

\newcommand{\idealworld}{%
\begin{tikzpicture}[baseline=0.1em,scale=0.9]
    \draw[semithick,rounded corners=0.8pt] (0,0) -- (0,0.15) -- (0.15,0.15) -- (0.15,0) -- (0,0);
    \draw[semithick,rounded corners=0.8pt] (0.3,0.3) -- (0.15,0.3) -- (0.15,0.15) -- (0.3,0.15) -- (0.3,0.3);
\end{tikzpicture}\xspace}

\begin{document}

\title[Effects of Code Duplication in Machine Learning Models of Code]{The Adverse Effects of Code Duplication\\in Machine Learning Models of Code}

\author{Miltiadis Allamanis}
\email{miallama@microsoft.com}
\affiliation{%
  \institution{Microsoft Research}
  \city{Cambridge, UK}
}

\begin{abstract}
  The field of big code relies on mining large corpora of code
  to perform some learning task towards creating better tools for
  software engineers. A significant threat to
  this approach was recently identified by \citet{lopes2017dejavu}
  who found a large amount of near-duplicate code on GitHub. However, the impact of
  code duplication has not been noticed by researchers devising
  machine learning models for source code. In this work, we explore
  the effects of code duplication on machine learning models showing that
  reported performance metrics are sometimes inflated by up to 100\% when testing
  on duplicated code corpora compared to the performance
  on de-duplicated corpora which more accurately represent how machine learning
  models of code are used by software engineers. We present a duplication index for widely used
  datasets, list best practices for collecting code corpora
  and evaluating machine learning models on them. Finally, we release tools
  to help the community avoid this problem in future research.
\end{abstract}



\maketitle

\section{Introduction}
Machine learning models of source code have
recently received great attention from the research community.
At the intersection of the research fields of software engineering,
programming languages, machine learning and natural language
processing, multiple communities have been brought together
into the field of ``Big Code'' or ``code naturalness''
with many fruitful results~\citep{allamanis2018survey}.
Commonly, research in this area relies on large corpora of code
which can be used as training and test sets, allowing machine learning
methods to learn and probabilistically reason about coding practice at a large scale.
The goal is to use the learned models to
provide useful tools to software engineers.

However, there is a looming crisis in this newly-founded area, caused
by a disproportionately large amount of code duplication. This issue ---
first observed by \citet{lopes2017dejavu} ---
refers to the fact that multiple file-level (near-)clones appear
in large corpora of code, such as those mined from GitHub
repositories. This is because software engineers often copy
--- partially or entirely --- files from other projects~\citep{lopes2017dejavu,gharehyazie2018cross}.
Despite the findings of \citet{lopes2017dejavu},
the research community has not yet investigated how and when code duplication 
negatively affects its research, the machine learning models it devises,
and the practical tools it creates. The core issue arises from the fact
that identical or highly similar files appear both in the training and
test sets that are used to train and evaluate the machine learning models.

In this work, we first describe the impact
that code duplication can have on machine learning models.
Although not all applications of machine learning models are affected
by code duplicates, a large majority of them is. 
We discuss the biases introduced when evaluating models under
duplication and show that duplication can cause the 
evaluation to overestimate the performance of a model
compared to the performance that actual users of the model observe.
Then, we replicate the work of
\citet{lopes2017dejavu} across ten corpora that have been
used in ``big code'' research and we
measure the impact of duplication across datasets and machine
learning models showing that the performance observed by
a user is up to 50\% worse compared to reported results.
Although this paper does not present any results or ideas that would
be unexpected to a statistician or a machine learning expert,
we hope that it will help programming language, software engineering and machine learning researchers
better understand the issue of code duplication for machine learning on 
code by clearly illustrating its impact. At the same time, we provide tools
and some best practices that can help overcome pitfalls when
researching machine learning methods that employ source code data.
We hope that this paper contributes the following:
\begin{squishlist}
  \item an application-driven principle for deciding if within the application domain
      code corpus deduplication is needed (\autoref{sec:duplicationAndML});
  \item the theoretical basis of the effects of code duplication (\autoref{sec:duplicationAndML})
     and a demonstration of the effects of code duplication on
     machine learning models of source code (\autoref{sec:impact});
  \item an \href{https://github.com/Microsoft/near-duplicate-code-detector}{open-source, cross-platform tool} that detects near-duplicates in
     C\#, Java, Python and JavaScript along with a \href{https://ieee-dataport.org/open-access/deduplication-index-big-code-datasets}{duplication index}
     for existing datasets, listing existing duplicate files (\autoref{sec:measuringDupl});
  \item a set of suggested best practices to mitigate
     the code duplication problem for machine learning models of code (\autoref{sec:discussion}).
\end{squishlist}

\section{Code Duplication \& Machine Learning}
\label{sec:duplicationAndML}
Code duplication refers to the idea that
a large snippet of code appears multiple times with no or small
differences within a corpus of code. Duplicates are a relatively small subset of
code clones~\citep{roy2007survey} --- a well-studied field of software engineering.
The existence of duplicates was noticed much earlier~\citep{tempero2010qualitas}
but their negative effect became significantly more noticeable due to recent advancements that
allowed the collection of large code corpora~\citep{lopes2017dejavu}.
In this paper, we are specifically interested in illustrating the effects of code duplication
on machine learning models of code\footnote{We use the terms ``duplicate'' and ``near-duplicate''
interchangeably to refer to code that is highly similar but not
necessarily identical. }.
This endeavor sets
different parameters for searching, understanding and classifying
code duplication. 
To understand the effects of duplicates, we first need to
discuss the practical applications of machine learning models for code.

Why do we want to train machine learning models on source code?
At a high-level, the goal
is to train models on existing code, such that the learned models
capture the statistical properties of some particular aspect of coding practice,
which can then be useful within a tool used by a software engineer.
Some examples of recently researched models include:
\begin{squishlist}
\item code completion models~\citep{hindle2012naturalness,raychev2014code,hellendoorn2017deep,maddison2014structured}
    aiming to assist code construction in an editor
    when a developer is writing new code. Such models are widely used in practice today.
\item Type prediction models~\citep{raychev2015predicting,hellendoorn2018deep}
    where the goal is to infer (or provide probabilistic hints for) the types
    of new, previously untyped, programs (\eg in JavaScript) ;
\item code summarization~\citep{allamanis2016convolutional,iyer2016summarizing,barone2017parallel,alon2018code2seq} where the
    goal is to summarize some code into a short natural language utterance.
\end{squishlist}
In most applications, like in the aforementioned examples, the goal is to use trained models to provide
recommendations and insights on \emph{new} and \emph{unseen}
code when the software engineer is creating or maintaining it. Essentially,
this necessitates that machine learning models
\emph{generalize} well to new source code or --- in statistical machine
learning terms --- to \emph{faithfully model the true distribution of
the data as it will be observed by the particular use case of the tool}.
As we will discuss later in this section, in order for a machine learning
model to generalize to the true data distribution, it needs to be trained
on data independently drawn from that distribution. Code duplicates commonly violate that.

Furthermore, the true data
distribution depends on the target application. Different applications
of machine learning models of code will tend to have different true
data distributions. Therefore, before training any machine learning
model of code, we should all ask \emph{``What is the 
distribution of the data that our machine learning component will need to operate on?''}

For example, for a token-level code completion model the true data distribution
refers to the predicted next token that the developer will actually type.
It is thus reasonable to assume that duplicate code is \emph{not} a
part of the true data distribution as a developer will copy-paste whole chunks rather than
type duplicate code character-by-character. However, there are other cases where code
duplication is part of the true data distribution. For example, if
we are interested in deobfuscating code that contains a lot of copy-pasted libraries/functions,
then duplicates are part of the true data distribution.

The duplication issue arises because, in practice, it is very rare for researchers to train their model
and measure its performance by directly observing its use by engineers,
\ie the true data distribution.
Instead, a common practice is to split any existing dataset into two parts:
a training set that is used to train the machine learning model
and a test set where the performance of the model is measured. 
And since duplicated datasets are distributed differently from
non-duplicated datasets the machine learning models learn to model
a different probability distribution. This is because machine learning
makes an important assumption: each of the data points need to be
\emph{independent and identically distributed} (i.i.d) over the true
distribution of data of the use case. This is \emph{not} an unreasonable assumption 
and is widely and successfully used in machine learning
and data mining research and practice~\citep[\S 7.3]{murphy2012machine}.
It is exactly this assumption that code duplication strongly violates
for many of the use cases of machine learning models of code.

In this paper, we make two assumptions.
First, the true data distribution
of the target application contains no duplicates. Second,
we assume that duplication happens only across
files, similar to \citet{lopes2017dejavu}. This means
that smaller amounts of code duplication, such as
clones that span only a few lines, are \emph{not} be considered duplicates.
The last assumption addresses the possibility that
the target use case of a machine learning-based software engineering tool
contains a few lines of cloned code. For example,
a type prediction tool may still be required to suggest
types even when a few lines of code have been copy-pasted.
These assumptions are central to the thesis of this paper:
As we will discuss later, particular use cases may allow for
duplicates within the true data distribution. The results presented
in this paper does not affect them. Other use cases may need
to consider additional type of duplicates, such as smaller 
cloned snippets or functional (type IV) clones. The results presented here are still
valid for those cases and, most probably, the negative effects of
code duplication would be more severe when a broader class of
code duplicates needs to be considered.

\begin{figure}\centering
    \includegraphics[scale=0.25]{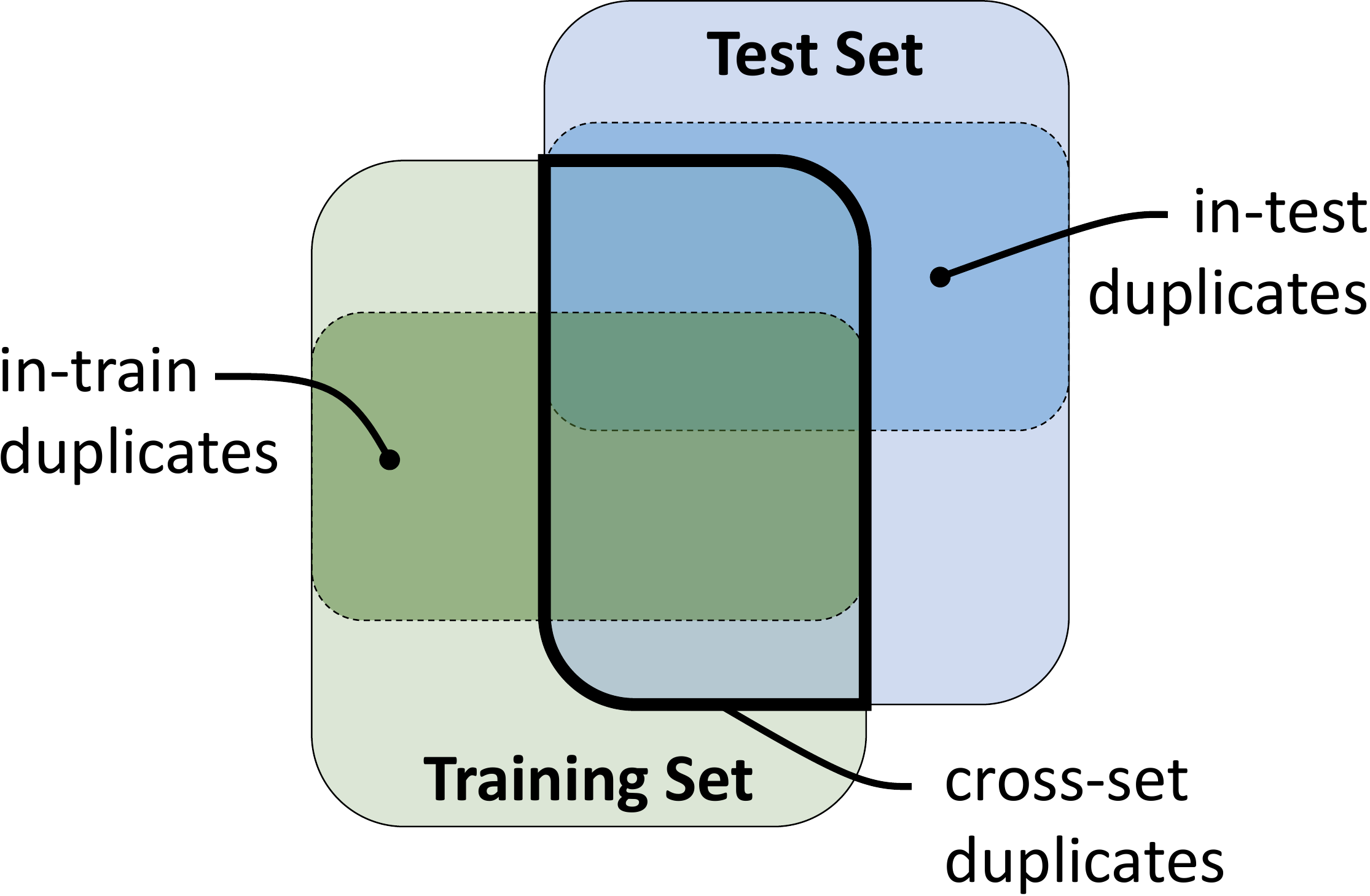}
    \caption{Schematic description of types of duplicates. The
    dashed boxes indicate the subset of files that are duplicates
    within each set.}\label{fig:conceptualDuplicateTypes}
\end{figure}

\paragraph{Concepts and Definitions}
Assume a dataset $D$ of source code files that is split into
a training and a test set (\autoref{fig:conceptualDuplicateTypes}).
We distinguish three types of duplicates: (1)
``in-train'' duplicates, \ie files
duplicated within the training set; (2) ``in-test''
duplicates, \ie duplicates within the test set; and (3)
``cross-set'' duplicates, \ie files that appear
both in the training and test sets.

\paragraph{Duplication Bias}
In machine learning, a measured quantity $f,$ such as the loss function
minimized during training or a performance (\eg accuracy) metric, is usually estimated as the average
of the metric computed uniformly over the training or test set(s)
(because of the i.i.d. hypothesis). Specifically, the estimate of $f$
over a dataset $D=\{x_i\}$ is computed as
\begin{align}\label{eq:festimator}
  \hat{f} = \frac{1}{\vert D \vert} \sum_{x_i \in D} f(x_i).
\end{align}
Duplication biases this estimate because some $x_i$ will appear multiple times.
Specifically, we can equivalently transform $D$ as a multiset
$X=\{(x_i, c_i)\}$ where $c_i \in \mathbb{N}_+$ is the number of times
that the sample $x_i$ is found in the dataset. Therefore, we can rewrite
\autoref{eq:festimator} as
\begin{align}\label{eq:duplicationf}
    \hat{f} = (1-d)  \underbrace{\frac{1}{\vert X \vert}\sum_{x_i \in X}f(x_i)}_{\text{unbiased estimate~}\bar{f}} + d\underbrace{\frac{1}{\vert D \vert - \vert X \vert}\sum_{x_i \in X}(c_i -1)f(x_i)}_{\text{duplication bias }\beta} 
\end{align}
where $d=\frac{\vert D \vert - \vert X \vert}{\vert D \vert}=\frac{\sum c_i - \vert X \vert}{\vert D \vert}$ is the \emph{duplication factor},
where $\vert X \vert$ is the number of unique $x_i$ in $X.$ Thus $d$ is the
proportion of the samples in the dataset that are duplicated ($c_i > 1$).
By rewriting the above equation as 
$\hat{f}=(1-d)\bar{f}+d \beta$
we see that the larger the duplication factor $d$, the larger
the effect of the duplication bias $\beta.$

From a machine learning perspective, the duplication bias in
the training loss causes a model to
overweight some training samples (the in-train duplicates).
During testing, the duplication
bias will skew the reported performance metric. Furthermore, we expect
cross-set duplicates to artificially improve any metric taking advantage
of the fact that multiple samples that are seen during training also appear in the test set,
giving the illusion that the model generalizes, 
where in fact it memorized duplicates.

\section{Measuring Duplication}
\label{sec:measuringDupl}
To measure code duplication we need a method that detects
(near) duplicate files along a large corpus of code. As we discussed
in the previous section, we are interested in file-level
duplication and thus we re-implement SourcererCC's~\citep{sajnani2016sourcerercc}
token-level duplication detection with minor modifications described 
next and release it under a permissive license.
These simple modifications adapt SourcererCC
to file-level duplicate detection, removing complexity that is
required for general-purpose code clone detection and are similar
to those discussed in \citet{lopes2017dejavu}.

\paragraph{Detecting near-duplicates} Although detecting exact duplicates is
straightforward, this misses a substantial number of near-exact
matches that differ only in a few aspects. To achieve this, we follow
SourcererCC~\citep{sajnani2016sourcerercc}: we tokenize each file
and extract all identifier and literal tokens. For each file, we build two
``fingerprints'', a
set $T_0$ and a multiset $T_1$ of all the identifiers and literals. We consider
two files $i$ and $j$ to be duplicates, if the Jaccard similarities $J(T^i_0, T^j_0)$
and $J(T^i_1, T^j_1)$ are above the thresholds $t_0$ and $t_1$ respectively.
In this work, we set $t_0=0.8$ and $t_1=0.7$
based on the default values used in SourcererCC and
experimentation on a C\# dataset, but we notice that
duplicate detection is fairly robust to these thresholds. Files
with fewer than 20 identifier tokens are not considered duplicates and are
excluded from our analysis. Finally,
to improve the speed of the tool, as in SourcererCC,
we make the simplifying assumption that similarity is transitive.
Although this does not generally hold, we found that this does not
impact the accuracy of the tool. Finally, since 
computing the Jaccard similarities is embarrassingly
parallel, we simply compare all combinations of files for similarity.

Our tool is quite fast. For example, on an Azure F16 machine
(2.4 GHz Intel Xeon E5-2673 v3 Haswell with 16 cores and the Intel Turbo Boost Technology 2.0 and 32GB of RAM), our 
method detects duplicates among 112k files in the JavaScript-150k
corpus (discussed next) in 5 hours.
We open-source the duplication-detection code online under
a permissive license at \url{https://github.com/Microsoft/near-duplicate-code-detector}. It contains
tokenizers for Java, JavaScript, C\# and Python but can easily
be extended to other languages. The deduplication tool
accepts a \href{http://jsonlines.org/}{JSONL} file (\ie a file containing a valid JSON per line) containing an id
of each file (\eg its filepath) and a list of identifier and literal
tokens within that file. It returns a JSON file with the
clusters of near-duplicate files. We also provide a faster, but approximate
Python tool that works on the same principles within the
\texttt{dpu-utils} package at \url{https://github.com/Microsoft/dpu-utils}.

\begin{table*}
    \caption{Duplication Statistics across Existing Corpora over all files (across any provided splits)
     with more than 20 identifier and literal tokens.} \label{tbl:duplicationStats}
    \centering\footnotesize
    \begin{tabular}{llrrrrrr} \toprule
Name & Relevant & \# Files      & \# Duplicate         & Duplicate & \multicolumn{2}{c}{Duplicate Group Size}& \% Expected Cross-Set Duplicate\\ \cline{6-7}
     &Publications &($\times 1000$)& Groups ($\times 1000$) & Files -- $d$ (\%) & Average  & Median &  Files within Test (6:4 split) \\ \midrule
C\#-19                 &\citep{allamanis2018learning}& 28.3  & 0.9    & 10.6 & 4.4 & 2  & 11.7 \\
Concode -- Java*       &\citep{iyer2018mapping}      & 229.3k& 30.8   & 68.7  & 6.1 & 3 & 77.8\\
Java GitHub Corpus     &\citep{allamanis2013mining}  & 1853.7  & 682.7  & 24.8 & 2.1 & 2 & 29.6\\
Java-Small             &\citep{alon2018code2seq}, \citep{allamanis2016convolutional}     & 79.8  &  2.4   & 4.7  & 2.6 & 2 & 5.7\\
Java-Large             &\citep{alon2018code2seq}     &1863.4 &195.0   & 20.2 & 2.9 & 2  & $^\dagger$24.1\\
JavaScript-150k        &\citep{raychev2016learning}  & 112.0 &  8.6   & 20.7 & 3.7 & 2  & 24.1\\
Python-150k            &\citep{raychev2016learning}  & 126.0 &  5.4   &  6.6 & 2.6 & 2  & 8.0\\
Python docstrings v1*  &\citep{barone2017parallel}   & 105.2 &  17.0  &  9.2 & 2.3 & 2  & 11.2\\
Python docstrings v2*  &\citep{barone2017parallel}   & 194.6 &  24.2  & 31.5 & 3.5 & 2  & 37.4\\
Python Autocomplete*   &\citep{hashimoto2018retrieve}&  70.4 &   8.9 &  20.3 & 2.6 & 2  & 24.5\\
        \bottomrule
    \end{tabular}

    {\scriptsize *We place one method per file, since the corpus is split across methods.
    $^\dagger$When the dataset is split across projects, as in the author provided split, this falls to 8.9\%.}
\end{table*}

\paragraph{Duplication Statistics}
Armed with a reasonable method for detecting duplication,
we now report code duplication statistics
for ten publicly available datasets that have
been used for machine learning on code. It should be noted
that for the studied datasets all authors have taken
significant steps to remove exact file-level clones. However,
this process missed a large number of (near) duplicate files,
that may differ in minor aspects, such as whitespace, code
comments and other small code modifications.
\autoref{tbl:duplicationStats}
reports the results. We note that for the JavaScript-150k
dataset our tool was able to process only 112k files\footnote{
    This is because the \href{http://esprima.org/}{\texttt{esprima}} parser failed to parse these files.
}
and therefore we report results on those files. The rest of the files
are ignored.
The results show that in many datasets, a substantial proportion of the dataset
contains duplicated code.
Note that these statistics are when datasets are split into different folds (chunks) across files.
When splitting across projects, this percent is most often reduced. For
example, splitting the Java-Large dataset across projects,
following the split provided by \citet{alon2018code2seq}, 8.9\% of the test
set is made of cross-set duplicates (compared to the average of 24.1\% when splitting
across files). This suggests that splitting across projects --- when possible --- is
a helpful strategy.


As expected,
smaller datasets, such as those collected over a small and curated set
of projects suffer less from duplication. The Concode dataset~\citep{iyer2018mapping}
seems to be the one suffering the most from
duplication, by having about 68.7\% of its methods be duplicates.
However, it should be appreciated that Concode and the Python docstring datasets are datasets
where each sample is a single function, rather than a full source code file.
If we transform the other datasets, such that each file contains a single function or a smaller snippet,
their duplication statistics might also
worsen. Note that once the data is split into training-test sets,
the percent of cross-set duplicates is smaller than the
full dataset duplication factor, since a noticeable proportion
of duplicates become in-train or in-test duplicates.
Finally, we note that the duplication in all datasets
is significantly smaller than that reported by \citet{lopes2017dejavu}.
This should be attributed to the fact that the corpus
collected by \citet{lopes2017dejavu} is orders of magnitude
larger than any of the datasets in \autoref{tbl:duplicationStats}.
Authors of the datasets discussed here made efforts to deduplicate and filter
the collected corpora by removing most low popularity projects
and some number of exactly duplicated files.
We release the duplicates files at
\url{https://ieee-dataport.org/open-access/deduplication-index-big-code-datasets}
We hope that these lists can be used as
dataset duplication index in future work.

\paragraph{Human Evaluation}
SourcererCC makes some approximations to make the search computationally efficient.
This raises the question about its precision.
The author of this paper inspected 100 random pairs
of duplicates for the Javascript-150k dataset~\citep{raychev2016learning}
and 100 random pairs from the Java-Large dataset~\citep{alon2018code2seq}
and annotated each pair as
a true or false positive. Overall, the duplicate detection achieves
perfect precision for both datasets. This is to be expected as
SourcererCC is a well-validated method and works very well for the special
and relatively easy case of detecting file-level duplicates.

Looking at the duplicates, we make a few qualitative, empirical observations. First,
we observe that a large majority of duplicates share the same file name.
For the JavaScript-150k, the majority of near-duplicates is of two kinds:
(a) different versions of the same file (b) configuration-like files that differ mostly on the
configuration values. In contrast, in the Java-Large dataset
we find more exact clones, duplicates of the same file but of
a different version and boilerplate code.
For the C\# corpus~\citep{allamanis2018learning}, we note that near-duplicates were mostly found within projects
and largely include autogenerated files. This is because the creator of
that dataset --- and author of this work --- had explicitly used
a similar process to check for and remove
duplicates when creating the dataset, but only across projects and under
stricter thresholds.

\section{Impact on Machine Learning Models}
\label{sec:impact}
So far, we have established that code duplication can --- in principle ---
have adverse effects to the way machine learning models of code
are trained and evaluated.  But is this
actually the case?
Analytically measuring the effect of duplication on machine learning
models in a generalized way is not possible.
This is because machine learning models differ widely
in their characteristics and we expect different
models and tasks to be affected differently by code duplication.
To empirically illustrate the impact of code duplication, we create
experimental settings that illuminate separate aspects
of the problem. In \autoref{subsec:biasVsUnbiased} and \autoref{subsec:capacity}
we focus on code autocompletion through language modeling.
This allows us to do an in-depth case study of a single model
and a few factors of variation. Then in \autoref{subsec:othermodels}
we train state-of-the-art models on other tasks. In all cases,
we assume a random 50-10-40 train-validation-test split
over the dataset. We use the validation set to evaluate training
decisions without exposing the model to the test set --- a standard practice
in machine learning.
For example, in neural networks where an algorithm iteratively
optimizes the model parameters, we pick the parameters for the
iteration that achieves the best performance on the validation set.
If a model does not use a validation
set, we merge the validation samples into the training set.

We note that this section does \emph{not} attempt to be exhaustive
but to replicate some recent work and study the
effects of duplication. Our goal is to merely elucidate
how these effects are demonstrated for the particular case of
machine learning models of source code, demonstrate that duplication
should \emph{not} be an afterthought when designing and evaluating
such models and help us distill meaningful best practices.

\begin{table}
    \caption{Terminology for Measuring Performance based on Kinds of Duplicates in Training and Test Sets}\label{tbl:perfTerms}
    \centering\footnotesize
    \begin{tabular}{llll} \toprule
    Training        & \multicolumn{3}{c}{Test Set}\\ \cmidrule(r){2-4}
                    & no dups & w/ cross-set dups & w/ all dups \\ \midrule
    Biased & Unbiased Test \userfacing & Cross-Set Biased \partrep & Fully Biased \reported\\
    Unbiased & Fully Unbiased \idealworld& -- & --\\
        \bottomrule
    \end{tabular}
\end{table}

\paragraph{Terminology} In the absence of existing terms, we
introduce a few new terms and annotate them with a
mnemonic symbol to help the reader.
Given a training-test split and by interpreting \autoref{eq:duplicationf},
we have two possible types of training:
\begin{squishlist}
\item \textbf{Unbiased Training} All duplicates are removed ($c_i=1, \forall i$ )
        and an unbiased loss function $\bar{f}$ is employed during training;
\item \textbf{Biased Training} All in-train duplicates are kept
        and the biased loss function $\hat{f}$ is used. Since most existing
        work does not adequately de-duplicate its datasets, it employs biased training.
\end{squishlist}
We now turn our attention to the testing terminology. Within
a testset we distinguish two types of duplicates: the
cross-set duplicates, and the in-test duplicates
(\autoref{fig:conceptualDuplicateTypes}). This leads
to four types of metrics, summarized in
\autoref{tbl:perfTerms} and discussed next. The mnemonic symbols can be
interpreted as Venn diagrams of the training and test sets.
When a set contains duplicates it is shaded (indicating bias on that set),
otherwise it is left blank. Finally, we
note that when we remove duplicates,
we keep exactly one file from each cluster of near-duplicates, such that
any duplicate file is used exactly once ($c_i=1$).
\begin{squishlist}
    \item \textbf{Fully Unbiased} \idealworld that represents
        an ``ideal world'', where all duplicates are removed both
        from training and test sets and the training and test sets are 
        completely disjoint, allowing us to perform
        unbiased training and testing.
    \item \textbf{Unbiased Test} \userfacing that represents
        the performance when the test set contains no duplicates.
        This is equivalent to the performance observed by a user who
        is using a machine learning model
        under the true data distribution, but the model has
        been trained in a biased way.
    \item \textbf{Cross-set Biased Test} \partrep which is
        the performance measured when performing a biased training
        and using a test set that only contains cross-set duplicates,
        but no in-test duplicates.
    \item \textbf{Fully Biased Test} \reported where training and
        testing happens on the duplicated (original) dataset. This is the
        metric that is reported by existing work. Compared to
        the cross-set biased test (\partrep) this metric is additionally
        biased by the in-test duplicates. Because this bias is arbitrary,
        it inhibits us from measuring the exact effect of code
        duplication. For this reason, we do \emph{not} report these metrics (\reported),
        but note that empirically it is always very close to the cross-set
        biased test metrics (\partrep).
\end{squishlist}
It should be noted that for estimating the impact of duplication on
machine learning models
it is technically incorrect to directly compare the
fully unbiased performance (\idealworld) with the unbiased test (\userfacing)
to measure the effect of code duplication. In contrast, comparison between
the cross-set biased (\partrep) and unbiased test (\userfacing) is
technically correct. This is
because when training a model on (slightly) different datasets, there
is no method that can distinguish between a model's capacity to learn
from more (but duplicated) data and the effect of duplication. In practice
we observe
negligible differences between deduplicated (\idealworld) and unbiased
testing (\userfacing) and we report both.

\subsection{Biased \vs Unbiased Performance}
\label{subsec:biasVsUnbiased}
As we discussed in \autoref{sec:duplicationAndML}, code duplication
can result in measuring better performance
compared to the one that a user would actually observe, negatively
impacting the user's experience.
In this and next section, we focus on the
effects of duplication on a single task, namely code autocompletion with language models.
By focusing on a single task and model we can do a deep-dive on various
aspects of code duplication and illustrate subtle effects.
Later, in \autoref{subsec:othermodels} we measure the impact of
code duplication on other models and on other tasks.

\paragraph{Autocompletion via Language Modeling} 
has been extensively studied 
both in natural language and in source code. The goal of
language models is to capture the statistical characteristics of
a language such that the output appears to be ``natural''.
Language models have been used for
autocompletion~\citep{hindle2012naturalness,raychev2014code,hellendoorn2017deep,maddison2014structured}
and it would be unreasonable
to assume that the true distribution of this particular use cases contains duplicate code.

To demonstrate the effects of code duplication we employ
a simple, yet powerful neural language model.
The goal is to show how even relatively
simple models are severely impacted by
duplication and draw observations that generalize to other models.
We follow the early work of \citet{bengio2003neural} for token-level
language modeling. Our neural language model (NLM) is described as
\begin{align}
    P(t_i) = \mathit{softmax}(E_o \sigma(W_c [E_i h(t_{i-1}) \dots E_i h(t_{i-c})]) + \mathbf{b})
\end{align}
where $E_o\in\mathbb{R}^{\vert V\vert\times K}$ and 
$E_i\in\mathbb{R}^{D\times\vert V\vert}$ are the output and input embedding matrices
of tokens, $W_c\in\mathbb{R}^{K\times c D}$ is a matrix, $\mathbf{b}$ is a bias
vector, and $h()$ is a function that takes a token and
converts it to a one-hot vector. All parameters are learned.
We train our model to minimize the empirical cross-entropy
on the training set, and pick the model that achieves the
best performance on the validation set. For simplicity,
in this work we set $K=D.$
Throughout this section, we set $D=128$,
train with RMSProp~\citep{tieleman2012lecture} and early stopping. As a vocabulary $V,$
we use the top 10k most frequent tokens. All results
are averaged across 5 runs on random splits of the data.

\newcommand{\stdev}[1]{{\scriptsize $\pm$#1}}

\begin{table}
\caption{Impact of Duplicates on Evaluation Performance on a simple Language Modeling Task 
on the reshuffled and slightly reduced JavaScript-150k \citep{raychev2016learning} dataset and standard deviations.}\label{tbl:userImpact}
\centering
\begin{tabular}{lrrrr} \toprule
       & \multicolumn{4}{c}{Performance}\\ \cmidrule(r){2-5}
Metric & \userfacing & \partrep & $\Delta(\userfacing,~\partrep)$ & \idealworld \\ \midrule
Acc (\%)    & 49.1\stdev{0.4}  & 55.1\stdev{0.4} & \color{red} -10.9\%   & 49.2\stdev{0.4}\\
Acc-ID (\%) & 8.6\stdev{0.7}   & 17.7\stdev{0.4} & \color{red} -51.4\%   & 8.3\stdev{0.3}\\
MRR         & 0.674\stdev{0.005} & 0.710\stdev{0.000}& \color{red} -5.1\%    & 0.674\stdev{0.005}\\
MRR-ID      & 0.136\stdev{0.005} & 0.224\stdev{0.005}& \color{red} -39.3\%   & 0.132\stdev{0.004}\\
PPL         & 9.4\stdev{1.0}   & 7.5\stdev{1.0}  & \color{red} +25.3\%   & 9.4\stdev{1.0}\\
PPL-ID      & 76.1\stdev{1.1}  & 55.4\stdev{1.1} & \color{red} +37.4\%   & 82.3\stdev{1.1}\\
\bottomrule
\end{tabular}
\end{table}

\paragraph{Performance}
To accurately measure the impact of duplication we need to be
able to make a fair comparison on the evaluated results.
To achieve this, we replicate the conditions of existing work, \ie
we perform biased training on our models. We then compute the
unbiased (\userfacing) and cross-set biased (\partrep) performance metrics.
\autoref{tbl:userImpact} shows the measured effect of duplication on
the reshuffled and slightly smaller JavaScript-150k dataset.
Specifically, it highlights the \% relative difference
between the unbiased-test (\userfacing) and cross-set biased (\partrep) metrics, which can directly
measure the effect of code duplication on the metrics. We also report
the fully-unbiased metrics (\idealworld). The metrics computed are
(a) the accuracy of
correctly predicting the next token (Acc; higher is better), (b) the mean reciprocal
rank (MRR; higher is better) over the tokens and (c) the perplexity (PPL; lower is better) assigned
by the neural language model. Unknown tokens are counted as
incorrect when computing accuracy and MRR.  We also compute focused
metrics on identifiers
since they have been proven to be the hardest to predict~\citep{allamanis2013mining,bielik2016phog,maddison2014structured}.
We note that we also computed the fully biased (\reported)
metrics and on average, the
NLM's performance is similar to
the cross-set biased (\partrep) performance. This is expected,
since the in-test bias is mostly random.

Based on the results, we notice that \emph{all}
metrics are affected to a different extent by code duplication.
The relative difference ($\Delta(\userfacing,~\partrep)$) ranges from a few percentage points to halved performance.
This suggests the seriousness of the code duplication problem.
Furthermore, we
observe that the identifier-related metrics are those that are
more severely affected by code duplication. This is expected,
since code duplication makes identifiers, which would otherwise
appear sparsely, appear more frequently and predictably.

Thus, it should be appreciated that \emph{not all metrics and tasks
are equally affected by code duplication.} For example, if an
application requires predicting code's non-identifier tokens (\eg as in \citet{campbell2014syntax}),
duplication would have a much smaller effect compared to an
autocompletion application for predicting identifiers.

\subsection{Model Capacity and Impact on Code Duplication}\label{subsec:capacity}
Duplication has an observable impact on
the performance of machine learning models of source code. However,
not all models are impacted in the same way. Indeed, some
models may be more prone to memorizing code duplicates than
others. Since we cannot directly compare the capacity of different models,
we perform a case study on the NLM model and illustrate how varying its
learning capacity causes the NLM to be affected differently by duplication.

\begin{figure}
    \includegraphics[width=\columnwidth]{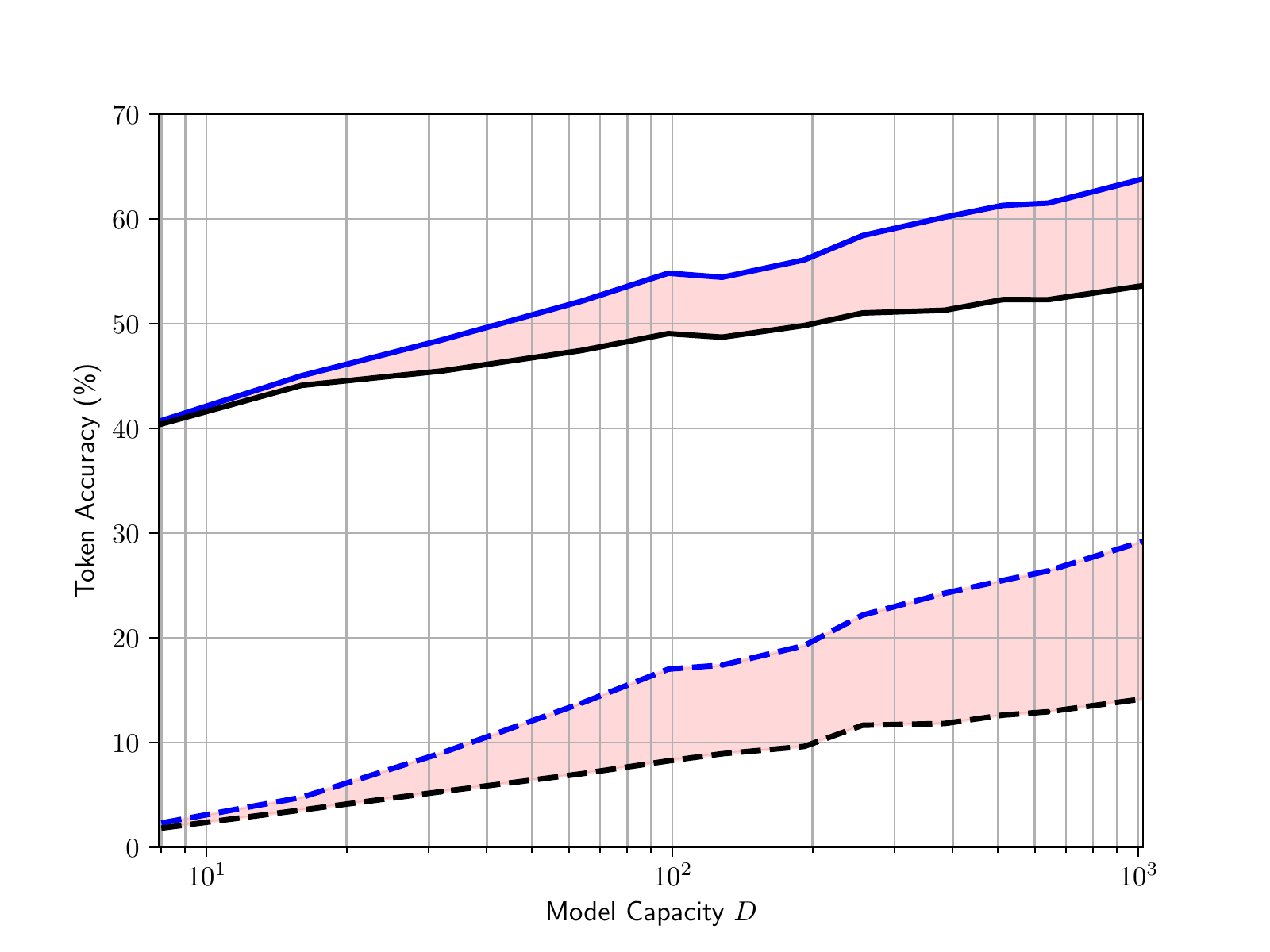}
    \caption{The impact of code duplication on the NLM with different
    capacity trained on JavaScript-150k. The solid lines show the accuracy of the NLM model
    when predicting all tokens, whereas the dashed lines show the accuracy
    of predicting only identifiers. Blue lines indicate the cross-set biased accuracy,
    and black ones show the unbiased test accuracy. The larger the capacity of the
    model, the more severe the impact of code duplication (red shaded area).} \label{fig:capacity}
\end{figure}

\autoref{fig:capacity} plots the NLM accuracy of predicting tokens (solid lines)
or only identifiers (dashed lines). As a proxy for measuring the capacity of the
model, we vary the dimensionality $D$ of the vector representations; a common
proxy for model capacity in the machine learning literature. Although there
are other methods to increase the capacity of the model (\eg by adding more
layers), increasing the dimensionality is a reasonable option for exploring the
effect of code duplication. The shaded (red) area
in \autoref{fig:capacity} shows, as expected, that the (negative) effect of duplication
increases as model capacity increases. This can be attributed to the fact that
additional capacity is used to memorize duplicated code. Therefore, we
observe that \emph{models that have larger capacity tend to be more heavily
affected by code duplication}.

This suggests an additional and important observation:
\emph{Comparison of different models under code duplication may not
be indicative of their real performance}. This is because some models,
having more capacity, can take better ``advantage'' of code duplication
and report improved results only because they are able to better
memorize the duplicated cross-set samples.

\subsection{Other Models and Tasks} \label{subsec:othermodels}
Previously, we illustrated the impact of code duplication
over a relatively simple neural language modeling task where we could control various
factors of variation and observe how different aspects of a model
are affected by code duplication. Although the reader probably
already suspects that code duplication affects many other models,
here we select a few state-of-the-art models and tasks to evaluate the impact
of code duplication. Again, note this is not an exhaustive evaluation, but
merely indicates how existing methods cope with code duplication on datasets
similar (and possibly reshuffled) to the ones used by the authors.
Our goal here is to illustrate the adverse effects of
duplication across a diverse set of models and tasks
where code duplication is \emph{not} part of the true
data distribution. It should be noted that none of the results presented here
should be interpreted as negative results for any of the existing
methods. Our study merely illustrates how different tasks
and state-of-the-art models are
also affected by code duplication. For example, the simple
neural language model of \autoref{subsec:biasVsUnbiased}
still has a significantly worse performance compared to PHOG (discussed next),
even after removing code duplicates.

\begin{table}
    \caption{Impact of Code Duplication on Performance over a Series of Methods/Tasks.
    $\Delta$ refers to the relative \% improvement (worsening). Note that some of 
    the evaluated methods are evaluated on different datasets compared to those
    used in the original works. }\label{tbl:othermethods}
    \centering
    \begin{tabular}{lrrrr} \toprule
           & \multicolumn{4}{c}{Performance}\\ \cmidrule(r){2-5}
    Metric & \userfacing & \partrep & $\Delta(\userfacing,~\partrep)$ & \idealworld \\ \midrule
    \multicolumn{5}{l}{\underline{Task}: Method Naming~~\underline{Model}: code2vec \citep{alon2018code2vec}} \\
    \multicolumn{5}{l}{\underline{Dataset}: Reshuffled Java-Large \citep{alon2018code2seq}} \\
    F1 (\%)             &  44.71        &  50.98        & \color{red} -12.3\%  &  46.04 \\
    Precision (\%)      &  53.00        &  58.92        & \color{red} -10.5\%  &  54.51 \\
    Recall  (\%)        &  38.67        &  44.93        & \color{red} -13.9\%  &  39.85\\ \\
    \multicolumn{5}{l}{\underline{Task}: Variable Naming~~\underline{Model}: \textsc{JsNice} \citep{raychev2015predicting}} \\
    \multicolumn{5}{l}{\underline{Dataset} : Reshuffled \& Reduced JavaScript-150k \citep{raychev2016learning}} \\
    Accuracy (\%)       &  34.44        &  55.04        & \color{red} -37.4\%  &  29.41\\ \\
    \multicolumn{5}{l}{\underline{Task}: Code Autocompletion \underline{Model}: PHOG \citep{bielik2016phog}} \\
    \multicolumn{5}{l}{\underline{Dataset} : Reshuffled \& Reduced JavaScript-150k \citep{raychev2016learning}} \\
    Accuracy (\%) -- Types               &71.80       & 75.69         & \color{red}  -5.1\%  &  72.95 \\
    Accuracy (\%) -- Values              &71.19       & 77.75         & \color{red}  -8.4\%  &  71.35\\
    ~~~~~~~~~~~~  -- Identifiers   &48.94       & 61.43         & \color{red} -20.3\%  &  49.05\\
    ~~~~~~~~~~~~  -- String Literal&25.62       & 43.89         & \color{red} -41.6\% &  24.51\\ \\

    \multicolumn{5}{l}{\underline{Task}: Docstring Prediction \underline{Model}: Seq2Seq \citep{barone2017parallel}} \\
    \multicolumn{5}{l}{\underline{Dataset}: Python Docstrings v1 \citep{barone2017parallel}} \\
    BLEU            & 12.32        & 13.86          & \color{red} -11.1\%  & ---\\
    \bottomrule
    \end{tabular}
\end{table}

\paragraph{Tasks and Models}
We select four reasonably well-known tasks
in the literature. 
Note that we re-split the datasets randomly assigning
each file to a set. This represents cases where
a model can be used within projects, which is often
a realistic scenario in machine learning-based software
engineering tools. Splitting across projects (as in the official Java-Large split), can
substantially reduce the impact of code duplication,
depending on the characteristics of each dataset.
\begin{squishlist}
    \item The \textbf{method naming} task of predicting the
        name of a method (function) given the body
        of the function (\ie summarization). Here we run the open-source 
        state-of-the-art code2vec model~\citep{alon2018code2vec} on
        the Java-Large corpus~\citep{alon2018code2seq}.
    \item \textbf{Variable Naming} which is the task of
        predicting the names of variables of a
        snippet of possibly obfuscated code. Note that
        we assume that the task is to deobfuscate new, previously
        unseen code rather than code whose deobfuscated form
        is known, as discussed in \citet{raychev2015predicting}.\footnote{This excludes some cases that the
        \textsc{JsNice} authors have observed in practice when they deployed
        it as a service. Specifically,
        in personal correspondence they mentioned to the author that submissions
        to the \textsc{JsNice} service often contain bundled parts of 
        various projects and libraries. As developers 
        use different versions of common libraries, \textsc{JsNice}
        needs to train/test on all the versions, not just one.
        
        The author of this work agrees with the \textsc{JsNice} authors.
        Indeed the application of deobfuscating code by matching it to (partially) previously
        seen code, requires training on duplicated data, since the duplicated dataset
        represents the true data distribution (\autoref{sec:duplicationAndML})
        of this partial ``soft-matching'' use case of \textsc{JsNice}.
        Thus, this particular use case is one where the true distribution contains
        duplicates.}
        We run
        the state-of-the-art non-neural \textsc{JsNice}
        model of \citet{raychev2015predicting} on the
        JavaScript-150k~\citep{raychev2016learning} dataset
        using the author-provided data extraction utility. Note
        that the split differs from the original one and some
        of the files are missing as discussed in \autoref{sec:measuringDupl}.
    \item \textbf{Code Autocompletion} which is the language modeling
        task used in the previous section. Instead of
        using the neural model of \autoref{subsec:biasVsUnbiased}, we employ the PHOG 
        model of \citet{bielik2016phog} another
        non-neural model. Since the code is not open-source
        yet, Pavol Bielik kindly helped with training
        and testing on that model. We provided the split on the
        reshuffled and slightly reduced
        JavaScript-150k~\citep{raychev2016learning} dataset
        for this task.
    \item \textbf{Documentation Prediction} which is the task
        of predicting the documentation (\eg docstring)
        of a function using its implementation.
        Here, the most recent approach is that of
        \citet{barone2017parallel} that use neural
        machine translation to ``translate'' code to
        documentation. Since the authors provided the output
        of their model, we use it directly to compute
        the performance, instead of performing
        our own training.
\end{squishlist}
Additionally, we considered the Variable Misuse task~\citep{allamanis2018learning}
which is the task of predicting
which type-correct, in-scope variable to use at a
given variable usage location. The only dataset
that is available here is that of \citet{allamanis2018learning}.
However, within the variable misuse sites only 0.5\% of the datapoints are duplicated. This
is due to the fact that the C\#-19 dataset~\citep{allamanis2018learning}
duplicates are mostly files that are semi-auto-generated,
such as assembly information files and resource files that
contain very few candidate variable misuse sites. Given
the duplication of 0.5\% we will \emph{not} consider this task.
Note that for all the tasks considered above, it would be
unreasonable to assume that the true distribution reflecting
the particular use case of each tool to contain any duplicates.
We train/test all these models with the default parameters
as provided by the authors in their open-source releases
of their code. 

\paragraph{Analysis of Results}
Overall, we observe in \autoref{tbl:othermethods} that
removing code duplicates noticeably reduces the measured performance
of all methods ($\Delta(\userfacing,~\partrep)$). Although all metrics
worsen, the effect differs. For example, JavaScript-150k
and Java-Large have very similar (file-level)
duplication but the impact of duplication on the evaluation metrics
of PHOG~\citep{bielik2016phog} and code2vec~\citep{alon2018code2vec} is quite different.
This can be attributed to two factors (a) different
models are affected differently (\eg because of their
inductive biases) (b) different tasks are affected
differently by code duplication.

An interesting observation is that training models with a biased dataset
(\userfacing) almost always results in worse performance compared to
training each model in an unbiased fashion (\eg without duplicates, \idealworld). This
may be due to the fact that part of each model's capacity
is spent on learning about duplicates, modeling a different data distribution and thus hindering the performance
of the model on the deduplicated test set. Thus, \emph{training on a biased
dataset usually has negative effects on model performance as observed
by end-users (\userfacing)}. \textsc{JsNice}, a non-neural method, seems to be
an exception. This may be attributed to the fact that
the reduced size of the deduplicated dataset harms
performance more than code duplicates due to the default hyperparameter values.
Finally, as we already observed, different metrics are
affected differently. A consistent theme has been
that identifier-related metrics (\eg accuracy of identifiers of PHOG
and of the NLM)
are the most severely impacted. Generalizing this, we can
conclude that this can be attributed to the sparsity~\citep{allamanis2018survey}
of some code constructs (\eg identifier names): \emph{Rare elements of code 
are hard to predict. Metrics and methods heavily relying on sparse constructs, such as identifiers,
are those most severely affected by code duplication}.

\section{Mitigating Duplication: Best Practices}
\label{sec:discussion}
In the previous sections, we believe that we were able
to document and sufficiently illustrate
the negative impact of code duplication on machine learning models of code.
We observed that:
\begin{squishlist}
  \item The target application of each machine learning model dictates
      whether duplicates need to be excluded from the training and testing data.
  \item Code duplication affects all metrics and the performance
     observed by end-users is often significantly worse than the one
     reported by evaluation metrics.
  \item Different metrics and applications are affected differently
     by code duplication.
  \item Powerful models that have larger capacity are
     impacted more by code duplication.
  \item Comparing different models using duplicated code corpora can be
     unfair to models with smaller capacity.
\end{squishlist}

\paragraph{Best Practices}
Through this paper, a set of best practices arise that we recommend to
researchers and practitioners:
\begin{squishlist}
\item \textbf{Understanding the True Data Distribution} for the target use-case.
    Does the distribution over which we expect the
    tool to be used contain duplicates? If not, then deduplication needs
    to be performed. If duplicates need to be removed, the granularity
    of duplicates should be considered. File-level duplication was studied
    in this work, but other use cases may require more or less fine-grained
    deduplication.

\item \textbf{Data Collection} Collecting large datasets in batch should
    be done carefully and deduplication methods --- like the one proposed
    by \citet{lopes2017dejavu} or the one used in this work\footnote{
       The tool can be found at \url{https://github.com/Microsoft/near-duplicate-code-detector} and an approximate version within the
       \texttt{dpu-utils} Python package at \url{https://github.com/Microsoft/dpu-utils}.} --- should be used
    to deduplicate the collected corpus. Simply removing exact matches and
    forks is a reasonable but clearly insufficient first step. Splitting
    the dataset across different projects, when possible, usually helps a lot, but
    duplication often still exists.

\item \textbf{Use of Existing Datasets} This work demonstrates varying
    levels of duplication for different datasets. However, duplication
    occurs to some extent in all existing datasets. When using existing
    datasets, we suggest using
    the duplication index provided in this work to remove duplicates.

\item \textbf{Model Capacity} Models that have a large capacity to memorize,
    suffer the most from the duplication problem and
    special attention should be given when evaluating them.
    Furthermore, researchers should include na{\"i}ve memorization
    methods in their baselines (\eg $k$ nearest neighbors). If these baselines perform
    ``too well'' compared to other widely-used models,
    this can indicate a duplication issue.
\end{squishlist}

Finally, it should be noted that while removing duplicates is often
the easiest option, small variations of (near) duplicates
may still be useful to learning more robust machine learning models.
An alternative to discarding duplicates is to down-weight duplicated
samples in the loss function and performance metrics, such that
each group of duplicated samples has the same weight as a single
deduplicated sample, \ie transform \autoref{eq:duplicationf} to
\begin{align}
    \bar{f} = \frac{1}{\vert X \vert}\sum_{x_i \in D}\frac{1}{c_i}f(x_i).
\end{align}

\paragraph{Other Considerations}
So far, we have considered the ``traditional'' option where a fixed
dataset is split for training and evaluation purposes. In some cases,
temporal data may be available, \eg the version history of a
codebase. Appropriately, slicing the dataset through time, training on older
code and testing on newer code, should be considered a valid evaluation
methodology. Nevertheless, code duplication still needs to be accounted.
For example, a developer might copy existing code and paste it into a new file,
thus ``contaminating'' a dataset with duplicates.

Similarly, deployment of machine learning models often necessitate that a model
is trained on the same codebase to the one where it operates on. Although this
may sound odd, the deployed machine learning model/tool will only observe
previously unseen code and therefore also operates on an unbiased test environment.
This emphasizes the divergence between an offline and an online evaluation
of some tool. In most cases, we are not able to perform online evaluation
of a model, which would provide the most accurate results. Instead offline
evaluations, common in academia and industry, should strive to replicate
the conditions of an online system.

\subsection{Conclusions}
We hope that this paper informs the research community about the
negative effects of code duplication on the evaluation of machine learning models and
informs practitioners about potential pitfalls when deploying such tools
in practice. Removing exact and near duplicates will allow for more accurate
comparison of machine learning models and methods and will lead
to better machine learning-based tools for programmers.

Finally, despite code duplication's negative effects many interesting
research opportunities arise. As \citet{kapser2008cloning} observe, code
clones are not always bad, as they often give developers additional
flexibility over the evolution of a project and, therefore, methods should embrace it. The work of
\citet{hashimoto2018retrieve} who combine retrieval methods
that find similar snippets within a database of code and then
perform edits over those examples is an interesting example
of such a direction.

Additionally, in contrast to most artifacts often studied in machine learning,
such as images and text, the independence assumption (i.i.d) may be too strong:
In contrast to common forms of data, code is created through an evolutionary,
incremental process. New software is created often because other code makes
the new software possible and new features often build up on functionality
that already exists. This evolution-like process of software,
implies a strong dependence between code that has been written
and code that will be written. On one hand, this enables ideas such as
big code and naturalness but at the same time complicates evaluation of
such ideas, as discussed in this paper. Researching machine learning models
and compatible programming language representations that can explicitly
take into account the correlations introduced by this
evolutionary process may allow for improved tools in this area.

Finally, code duplication across code is a fact of software engineering life
and interesting
research questions such as ``Can new machine learning tools be
created that are robust to code duplication?'' and ``Can we usefully
exploit near-duplicates to produce better software engineering tools?''
seem to arise as interesting research problems.

\begin{acks}
  The author would like to thank Marc Brockschmidt for useful discussions
  and suggesting the mnemonic symbols,
  Patrick Fernandes for first noticing the severity of the duplication problem
  and bringing it to the attention of the author and
  an anonymous reviewer of some other work of the author that insisted that code duplication is not
  an important issue in existing datasets.
  Finally, the author would
  like to thank Pavol Bielik for running the evaluation on PHOG,
  Uri Alon for useful discussions on the Java-Large corpus and useful
  comments on a draft of this work, Charles Sutton and Earl Barr for helpful
  discussions, suggestions and corrections
  and anonymous reviewers and SPLASH Onward! PC for helpful comments and suggestions.
\end{acks}

\bibliographystyle{ACM-Reference-Format}
\bibliography{bibliography}

\end{document}